\documentclass[twocolumn,showpacs,preprintnumbers,amsmath,amssymb,prl]{revtex4}

\usepackage{graphicx}
\usepackage{dcolumn}
\usepackage{bm}

\begin{document}

\preprint{APS/123-QED}

\title{Phase Coherence Effects in the Vortex Transport Entropy}

\author{G. Bridoux}
\email{bridoux@cabbat1.cnea.gov.ar}
\author{G. Nieva}

\author{F. de la Cruz}
\affiliation{Centro At\'omico Bariloche and Instituto Balseiro,
Comisi\'on Nacional de Energ\'{i}a At\'omica, Av. E. Bustillo
9500, R84002AGP S. C. de Bariloche, Argentina}


\date{\today}

\begin{abstract}
Nernst and electrical resistivity measurements in superconducting
YBa$_2$Cu$_3$O$_{7-\delta}$ (YBCO) and
Bi$_2$Sr$_2$CaCu$_2$O$_{8+\delta}$(BSCCO) with and without
columnar defects show a distinctive thermodynamics of the
respective liquid vortex matter. At a field dependent high
temperature region in the $H-T$ phase diagram the Nernst signal is
independent of structural defects in both materials. At lower
temperatures, in YBCO, defects contribute only to the vortex
mobility and the transport entropy is that of a system of vortex
lines. The transition to lower temperatures in BSCCO has a
different origin, the maximum in the Nernst signal when decreasing
temperature is not associated with transport properties but with
 the entropy behavior of pancake vortices in the presence of
structural defects.

\end{abstract}

\maketitle

 The Nernst signal in superconductors is associated with
the displacement of vortices induced by the presence of a
temperature gradient, \textbf{$\nabla$}$T$, perpendicular to the
internal magnetic field, \textbf{B}. Its detection is made
measuring the Josephson voltage induced by vortices crossing
electrical contacts aligned in a direction perpendicular to both,
\textbf{B} and \textbf{$\nabla$}$T$.  For an ideal superconductor
and straight vortex lines if \textbf{$\nabla$}$T$ is in the
\textbf{\^{x}} direction, (\textbf{$\nabla$}$T$)$_x$, and
\textbf{B} in the \textbf{\^{z}} direction the Nernst electric
field is proportional to (\textbf{$\nabla$}$T$)$_x$
\cite{CampbellEvets}. Within this regime it was shown
\cite{CampbellEvets} that the thermal force per unit vortex length
is $F_T(H)= S_{\phi}(H,T)(\textbf{$\nabla$}$T$)_x $. Here
$S_{\phi}(H,T)$ is the transport entropy per unit vortex length.
In this limit the impedance to vortex displacement is the flux
flow vortex viscosity, associated with the flux flow electrical
resistivity, $\rho_f$. Thus, the Nernst signal is found \cite{Ri}
\begin{equation}
\label{eqen} e_{N}(H,T) = \frac {E_y} {(\nabla T)_x} = \frac
{\rho_f S_{\phi}(H,T)}{\phi_0}
\end{equation}
where $E_y$ is the Nernst electrical field and $\phi_0$ the flux
quantum. In the normal state $e_N(T,H)$ is essentially zero
\cite{Ong} and  since $S_{\phi}(H,T)=0$ at either $T=0$ or $H$
less than the lower critical field $H_{c1}(T)$ we see that $e_{N}$
should show a maximum as a function of $H$ or $T$.

In real materials vortex pinning inhibits vortex displacements
within an ohmic regime \cite{FisherErnesto}. Therefore, the
maximum of $e_N(T,H)$ is usually determined by the field or
temperature where pinning reduces vortex mobility to zero. In low
$T_c$ materials this imposes strong limitations on the range of
fields and temperatures where $e_N(T,H)$ can be used to determine
$S_{\phi}(H,T)$. In high $T_c$ superconductors the short coherence
length, strong material anisotropy and thermal energy are
responsible for a transition from a solid to a liquid vortex state
in a wide region of the $H-T$ phase diagram. This phase diagram
for a superconductor opened a renewed interest in the study of
vortex physics: solid-liquid phase transitions, vortex cutting and
reconnection, low vortex dimensionality, vortex decoupling and the
contribution of thermal fluctuations play an important roll
\cite{Blatter}.

The pioneering work by Ri et al. \cite{Ri} showed that the Nernst
effect in films of YBCO and BSCCO responds, qualitatively, to the
behavior expected from Ginzburg-Landau theory. Recent work by Ong
and collaborators \cite{Ong} triggered intensive experimental and
theoretical activity.

Previous work on YBCO and BSCCO showed \cite{Phylos}
\cite{SafarBush} that using the dc flux transformer contact
configuration in twinned YBCO crystals the $c$ axis vortex phase
correlation across the sample was established at a well defined
sample thickness dependent temperature, $T_{th}(d,H)$. At
$T_{th}(d,H)$  the resistivity in the $c$ direction,
$\rho_{c}(H,T)$,
 drops abruptly and the resistivity in the $ab$ plane, $\rho_{ab}(H,T)$,
 of phase correlated vortices across the sample decreases \cite{Phylos}
 and becomes zero at a thickness independent temperature $T_i(H)$, where the
  vortex liquid-solid transition takes place.
In BSCCO no vortex phase correlation \cite{SafarBush} across the
sample in the $c$ direction was established in the liquid region
of the $H-T$ phase diagram. From these transport properties of the
two paramount high $T_c$ superconductors we expect that if the
maximum of the Nernst signal is due to the decrease of  vortex
mobility it should take place at the proximity of $T_{th} (d,H)$
in YBCO and of $T_i(H)$ in BSCCO.

To study the relevance of the vortex phase correlation in the
field direction we measured $e_N(T,H)$ and the resistivity of
optimally doped single crystals of YBCO and BSCCO with and without
columnar defects, CD, for $\textbf{B}\|\textbf{c}$ axis. In
twinned YBCO crystals the sample thickness dependent flux cutting
and reconnection induces the field and temperature dependent
maximum of $e_N(T,H)$ but it does not contribute to the vortex
transport entropy. We show that the maximum in $e_N(T,H)$ in YBCO
below $T_c$ is induced by the size dependent vortex mobility.
While in YBCO the vortices in the liquid state are accepted to
respond as a three dimensional system, in BSCCO the vortices are
considered to be \cite{BulaKosh} uncoupled pancakes nucleated on
Cu-O planes. We have found that the maximum in the Nernst voltage
below $T_c$ in BSCCO is not associated with the vortex mobility in
the liquid state but to an intrinsic change of the temperature
dependence of $S_{\phi}$ of vortices in this phase.

The YBCO and BSCCO single crystals were grown as described in
references \cite{Phylos} and \cite{Kaul}. The columnar defects,
nearly parallel to the $c$ axis, with a dose equivalent field of
$B_{\phi}=3\,$T were created by irradiation with 278 MeV
Sb$^{24+}$ for BSCCO and 309 MeV Au$^{26+}$ ions for YBCO at
TANDAR-Argentina. The ions tracks are of the order of the
coherence length of both materials (5-10 nm in diameter).
 The $T_c$'s of the non-irradiated samples used in the Nernst and the electrical
resistance experiment were $T_c =$ 91.0 K and 91.3 K respectively
for BSCCO and 93.6 K for YBCO. The irradiated samples have $T_c =$
90.2 K for BSCCO and $T_c =$ 91.7 K for YBCO. The samples
thickness were 20-40 ${\mu}$m for the non-irradiated samples and
8-15 ${\mu}$m for the irradiated ones. The Nernst \cite{angulo}
and the electrical resistance \cite{Phylos} measurement setup were
described elsewhere.

\begin{figure}[h]
\includegraphics[width=6.8cm]{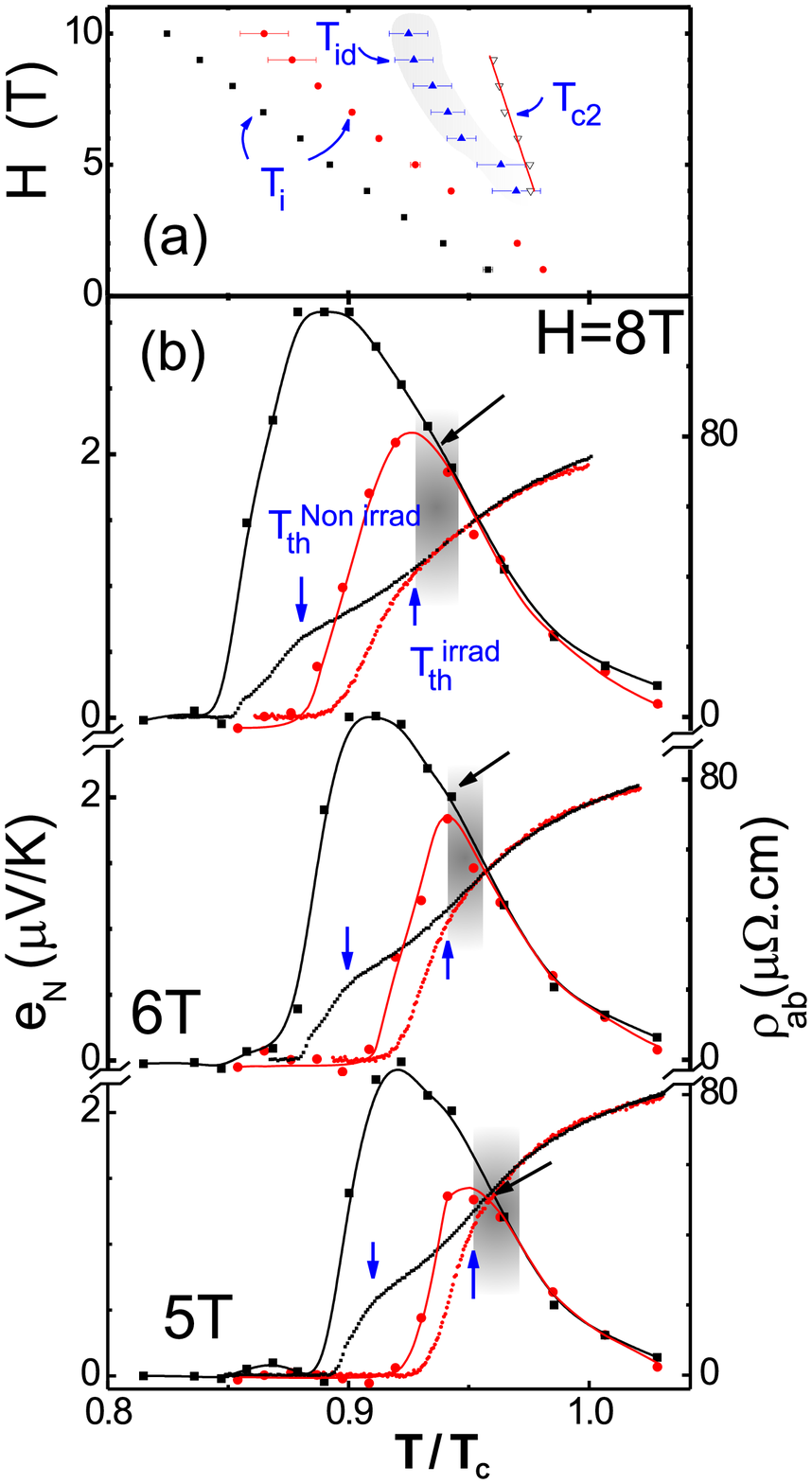}
\caption{\label{Fig1}
(a) $H-T$ diagram. The lines $T_{id}$ (up triangles), $T_{c2}$
(down triangles and linear fit) and the irreversibility lines
$T_i$ for the irradiated (circles) and non-irradiated samples
(squares) are shown. (b) Nernst signal $e_N$ and electrical
resistivity $\rho_{ab}$ vs. $T/T_c$ in YBCO at $H=5,\,6$ and
$8\,$T for the irradiated (circles) and non -irradiated samples
(squares). $T_{th}$ of both samples are indicated with arrows
following the references \cite{Phylos} \cite{Righi}. Gray regions
in both panels are equivalent.}
\end{figure}

In Fig.\ref{Fig1}(b) we show $e_N(T,H)$ and the electrical
resistivity of YBCO crystals for $H= 5$, 6 and 8 T with and
without CD. Once the vortex system melts the mobility is finite,
$e_N(T,H)$ grows fast with temperature up to a maximum at
$T_{max}(H)$. At higher temperatures the Nernst voltage decreases
towards the almost zero normal state value at temperatures well
above $T_c$. The mobility edge of the irradiated sample moves to
higher $T$ and so does the corresponding $T_{max}(H)$. The data
suggest that $e_N(T,H)$ between the mobility edge and $T_{max}(H)$
is strongly determined by pinning.

More important, $e_N(T,H)$ of samples with and without CD
coincides, within experimental uncertainty, at temperatures equal
and above $T_{id}(H)$ as shown by a tilted arrow in
Fig.\ref{Fig1}(b). It is found that the resistivities of the
samples with and without CD become equal in the same range of
temperatures, see Fig.\ref{Fig1}(b). This is observed in the whole
range of field investigated, from 2 to 10 T, suggesting that for
$T \geq T_{id}(H)$ the columnar defects and, possibly, all other
type of pinning centers, are irrelevant for the vortex response to
gradient of temperatures or electrical currents. Thus,
$\rho_{ab}$, $e_N$ and consequently the entropy of samples with
and without CD coincide for $T \geq T_{id}(H)$.

From the previous discussion and experimental observation we
define $T_{id}(H)$ as the highest temperature where vortex pinning
is effective, as determined by $e_N$ measurements. Above
$T_{id}(H)$ the pinning potential becomes irrelevant,
independently of  the type or strength of the pinning potential.
As we demonstrate below for $T < T_{id}(H)$ not only the pinning
potential of the columnar defects is switched on but also that
associated with the defects in the non-irradiated sample as made
evident by  the behavior of $\rho_{ab}(H,T)$, see
Fig.\ref{Fig1}(b), and the corresponding transport entropy of the
vortices. This makes $T_{id}(H)$  a relevant line in the phase
diagram of the liquid state of vortex lines. The gray region in
Fig.\ref{Fig1}(b) indicates the experimental uncertainty for
$T_{id}(H)$. In Fig.\ref{Fig1}(a) the line $T_{id}(H)$ is plotted
together with the pinning potential dependent solid-liquid phase
transition lines $T_i(H)$.

From the definition of $T_{id}(H)$ we recognize that for $T\geq
T_{id}(H)$, $U_{\phi}(T,H)= TS_{\phi}(H,T)$ should be the same for
the YBCO samples investigated, independently of the type of
pinning potential characterizing the transport properties at lower
temperatures. Therefore,  $U_{\phi}(T,H)$  becomes an intrinsic
property of the ideal vortex system for  $T \geq T_{id}(H)$.

\begin{figure}[h]
\includegraphics[width=8cm]{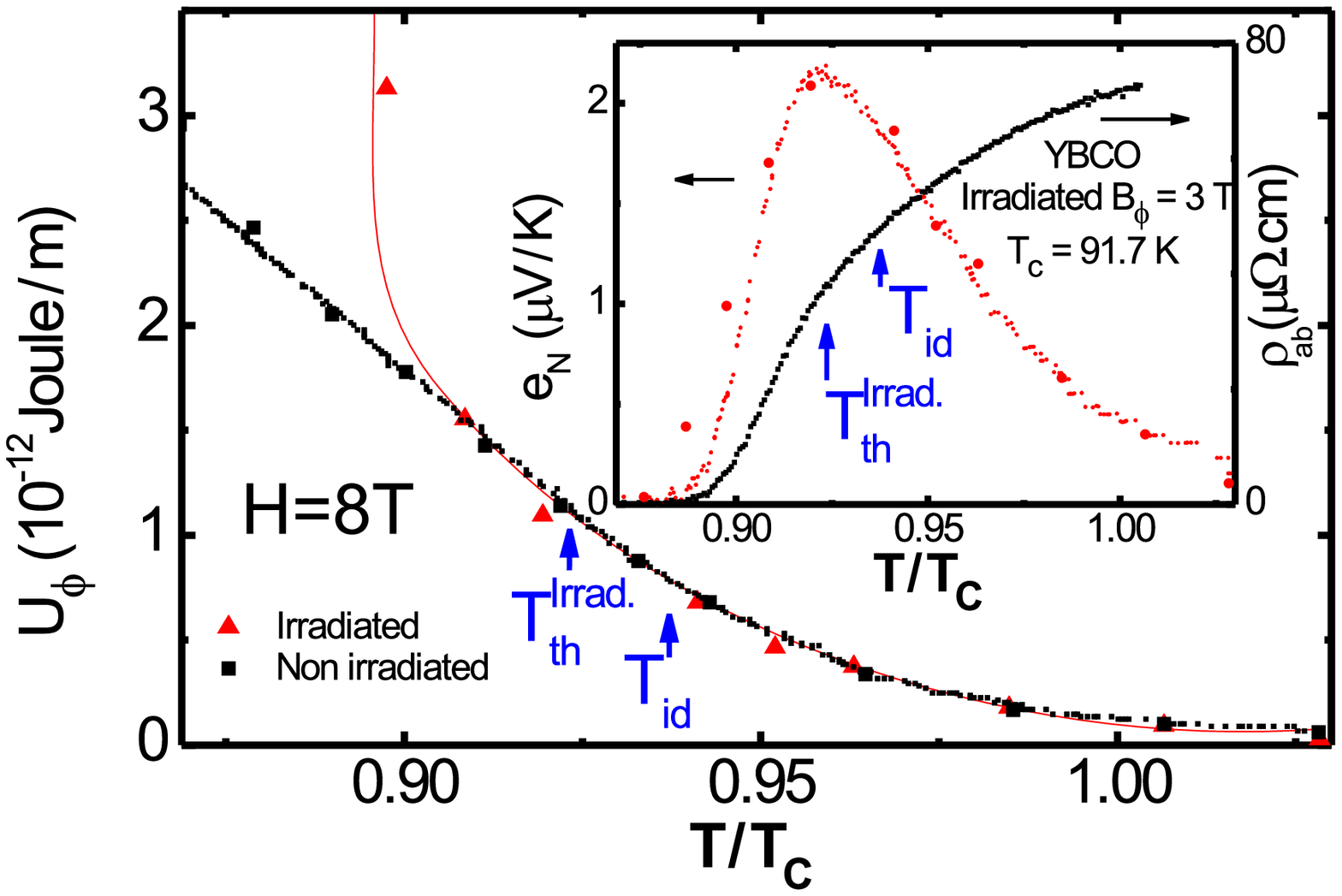}
\caption{\label{Fig3}
 U$_{\phi}$ vs.  $T/T_c$ in YBCO at $H=8\,$T
 for the irradiated (red triangles plus line) and the non-irradiated
  sample (black squares, interpolation with small symbols). Inset:
  Measured $e_N$  and $\rho_{ab}$ (red circles and black squares respectively)
   and recalculated $e_N$ (small red circles) for the irradiated
    crystal.}
\end{figure}

In Fig.\ref{Fig3} we plotted $U_{\phi}(T,H)$ of samples with and
without CD as a function of temperature at 8 T. The data are
representative of  $U_{\phi}(T,H)$ in the whole range of fields
investigated. We see that both $U_{\phi}(T,H)$ coincide not only
for $T \geq T_{id}(H)$ but also in a broader range of temperatures
below $T_{id}(H)$. This demonstrates that the maximum of
$e_N(H,T)$ is only induced by the effective pinning potential
acting on electrical transport properties, see inset in
Fig.\ref{Fig3} for the results in the irradiated sample. Somewhat
below its $T_{th}(H)$, $U_{\phi}(T,H)$ shows an anomalous increase
when decreasing temperature as compared to that of the
non-irradiated sample. The analysis of this feature helped to
understand the relative contribution of $\rho_{ab}$ and $S_{\phi}$
to $e_N$ in YBCO.

We see from Fig.\ref{Fig3} that $U_{\phi}(T,H)$ of the
non-irradiated sample increases linearly when decreasing $T$ well
below the $T_{th}(H)$ of the irradiated sample. This linear
dependence is expected in a mean field description
\cite{CampbellEvets}, its extrapolation to $U_{\phi}(T,H) =0$
provides a mean field $T_{c2}(H)$, also plotted in
Fig.\ref{Fig1}(a). The results support the idea that the measured
$U_{\phi}(T,H)$ is that of an intrinsic bulk superconductor,
independently of the amount and type of pinning centers. Assuming
this we calculated the $e_N(H,T)$ of the sample with CD using its
measured $\rho_{ab}$ and the $U_{\phi}(H,T)$ from the sample
without CD, as plotted in the inset of Fig.\ref{Fig3}. The results
are important and revealing. We see that the calculated $e_N(H,T)$
reproduces the experimental data within the experimental error
from temperatures well above $T_c$ to somewhat below the
temperature at the maximum. More important, the calculated data
extrapolates to the same $T_i(H)$, detected by electrical
transport. This result makes evident the origin of the anomalous
raise of $U_{\phi}(T,H)$ at temperatures below the maximum, in
both samples. The thermal gradient used to determine the Nernst
effect induces a force that exceeds the extremely small linear
response regime close to the solid-liquid transition
\cite{Grigera}. The thermal force in this range of temperatures is
calculated to be \cite{Bridoux} up to one order of magnitude
larger than that used in electrical measurements. The results show
that the maximum of $e_N(H,T)$ in YBCO is a size effect that
reflects the growth of the vortex phase coherence in the $c$
direction. The maximum takes place close to $T_{th}(d,H)$ where
the $c$ axis vortex phase correlation length coincides with the
sample thickness.

\begin{figure}[h]
\includegraphics[width=6.5cm]{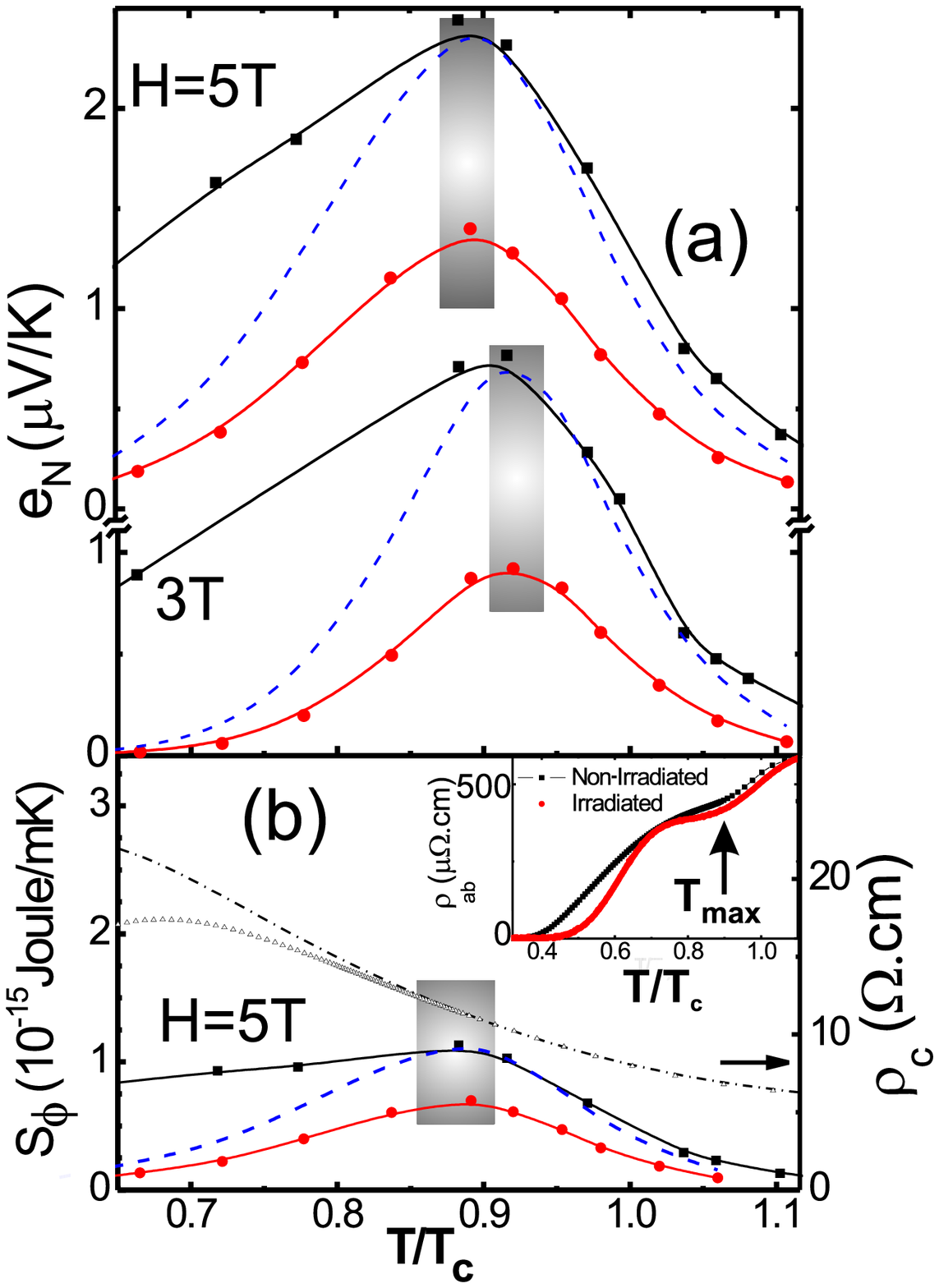}
\caption{\label{Fig4}
 (a) Nernst signal $e_N$ at $H= 3$ and
  $5\,$T vs. $T/T_c$ in BSCCO . Red circles
 and black squares curves correspond to the irradiated
 ($B_{\phi}=3\,$T) and non-irradiated sample respectively.
  The dashed curves correspond to the
Nernst signal of the irradiated sample normalized at the
corresponding maximum of the non-irradiated one. The shading
emphasizes the maximum region. (b) Entropy per unit vortex length
$S_{\phi}$ ($S_{\phi}= e_N{\phi}_0/\rho_{ab}$) vs. $T/T_c$ in
BSCCO at $H=5\,$T for the irradiated (red circles) and the
non-irradiated samples (black squares). The dashed curve
corresponds to $S_{\phi}$ of the irradiated sample normalized at
the corresponding maximum of the non-irradiated one. The $\rho_c$
vs. $T/T_c$ at $H=5\,$T (up triangles) and $8\,$T (dash-dotted
line). The inset shows $\rho_{ab}$ at $H=5\,$T for the samples
with (red circles) and without CD (black squares).}
\end{figure}

While the maximum occurs at a thickness dependent temperature the
inflection of the Nernst signal occurs at the thickness
independent $T_{id}(H)$. A remarkable result associated with
$T_{id}(H)$ is that in this line the linear temperature dependence
characteristic of a mean field behavior of $U_{\phi}$ is lost, see
Fig.\ref{Fig3}. At higher temperatures the curvature of
$U_{\phi}(T,H)$ indicates the dominant contribution of thermal
fluctuations, insensitive to the presence of structural defects.
In this sense $T_{id}(H)$ might well be an experimental
verification of that proposed by Nguyen and Sudb\o \cite{Subdo}
where thermally induced vortex loops proliferation causes the loss
of vortex line tension.

The Nernst signal and electrical resistivity of BSCCO samples with
and without CD for fields from 1T to 16 T were measured.
Fig.\ref{Fig4}(a) shows typical results, in this case for $3$ and
$5\,$T. In BSCCO we found that the absolute value of $e_N(H,T)$ of
the sample with CD is systematically smaller than that of the
non-irradiated sample. It is remarkable that for $T \geq
T_{max}(H)= T_{id}(H)$ the ratio of $e_N(H,T)$ between the
irradiated and non-irradiated samples is only a function of $H$
and not of temperature \cite{befi}, see Fig. \ref{Fig4}(a).

  At first glance some similarities are found between the results of BSCCO and YBCO:
  For  $T \geq T_{id}(H)$, the $e_N(H,T)$ of YBCO as well as
  the normalized $e_N(H,T)$ of BSCCO in samples with and without CD coincide up to
temperatures well above their respective $T_c$; for $T \leq
T_{max}(H)$, $e_N(H,T)$ goes to zero at the corresponding $T_i(H)$
in each material. On the other hand differences are evident.
Contrary to what is measured in YBCO, $T_{max}(H)$  of BSCCO is
the same for samples with and without CD, as well as for samples
of different thicknesses \cite{Ri}; in YBCO the rapid decrease of
$\rho_{ab}(H,T)$ in each type of sample determines the maximum of
$e_N(H,T)$; in BSCCO $\rho_{ab}(H,T)$ is independent of the
presence of CD to temperatures well below $T_{max}(H)$, see inset
of Fig. \ref{Fig4}(b). The analysis of $e_N(T,H)$ and
$\rho_{ab}(H,T)$ for samples with and without CD allows the
detection of magnitudes associated with the equilibrium
thermodynamic state of the vortex system. In YBCO, $T_{max}(H)$ is
determined by the behavior of $\rho_{ab}(H,T)$ approaching
$T_i(H)$. Thus, the maximum  of $e_N(H,T)$ is not associated with
an equilibrium state. On the contrary, that maximum in BSCCO is
determined by the maximum of an equilibrium property
$S_{\phi}(H,T)$, Fig.\ref{Fig4}(b), for all fields and
temperatures studied \cite{Bridoux}. This is an interesting and
  puzzling result: While the vortex mobility,  $\rho_{ab}(H,T)$ as well as the
  temperature $T_{max}(H)= T_{id}(H)$ remain unaffected by the nature of the
  pinning potential for temperatures in the neighborhood of the maximum
  of $e_N(H,T)$, it is the thermodynamic equilibrium property
  $S_{\phi}(H,T)$ that becomes dependent on the type of structural defects for
$T<T_{id}(H)$.
 The liquid vortex state in BSCCO is accepted to
be \cite{BulaKosh} that of 2D uncoupled vortices nucleated in Cu-O
planes. On the other hand, the data in this work show that below
the temperature  $T_{id}(H)$ (identical for samples with and
without CD) $S_{\phi}(H,T)$ becomes dependent on the type of
structural defects. Thus, the order induced by correlated defects
below $T_{id}(H)$ requires a change of the nature of the vortex
system at this temperature. This is supported by the incipient
decrease of $\rho_c(T,H)$ from the non-metallic normal state at
$T_{id}(H)$ in both samples, see Fig. \ref{Fig4}(b). This points
out that for $T< T_{id}(H)$ the interaction between pancakes
vortices in the neighbor Cu-O planes should not be disregarded.
This interaction could induce a change of the effective
dimensionality of the vortex system in BSCCO, decreasing the
configurational entropy contribution \cite{Huse} in the presence
of columnar defects.

In conclusion, transport and Nernst measurements in samples with
and without correlated defects reveal the origin of the maximum of
the Nernst signal in the liquid state of dissimilar vortices. In
YBCO, vortex cutting and reconnection \cite{Blatter} induces a
sample size dependent $T_{max}(H)$, well below the corresponding
$T_{id}(H)$. In this case the  vortex transport entropy remains
independent of the nature of defects. On the contrary, in BSCCO
the maximum is due to the intrinsic maximum of the vortex
transport entropy, suggesting a change of the effective vortex
dimensionality below the well defined temperature where the
resistance in the $c$ direction starts to deviate from its  normal
state value.

We acknowledge correspondence with L. Bulaevskii and D. Huse. We
thank L. Civale and H. Lanza for the irradiation, E. E. Kaul and
N. Saenger for help with sample preparation and measurements. G.
B. and G. N. acknowledge financial support from CONICET-Argentina.

\end{document}